# Specific heat, magnetic susceptibility, resistivity and thermal expansion of the superconductor ZrB$_{12}$

R. Lortz[1], Y. Wang[1], S. Abe[1], C. Meingast[2], Yu. B. Paderno[3], V. Filippov[3], and A. Junod[1,‡].

[1] Department of Condensed Matter Physics, University of Geneva, CH-1211 Geneva 4 (Switzerland).
[2] Forschungszentrum Karlsruhe, Institut für Festkörperphysik, 76021 Karlsruhe, Germany.
[3] Institute for Problems of Materials Science NANU, Kiev, Ukraine.

**Abstract**

In an attempt to clarify conflicting published data, we report new measurements of specific heat, resistivity, magnetic susceptibility, and thermal expansivity up to room temperature for the 6 K superconductor ZrB$_{12}$, using well-characterized single crystals with a residual resistivity ratio >9. The specific heat gives the bulk result $2\Delta(0)/k_BT_c = 3.7$ for the superconducting gap ratio, and excludes multiple gaps and *d*-wave symmetry for the Cooper pairs. The Sommerfeld constant $\gamma_n$ = 0.34 mJ K$^{-2}$ gat$^{-1}$ and the magnetic susceptibility $\chi$ = $-2.1 \times 10^{-5}$ indicate a low density of states at the Fermi level. The Debye temperature $\theta_D$ is in the range 1000-1200 K near zero and room temperature, but decreases by a factor of ~2 at ~35 K. The specific heat and resistivity curves are inverted to yield approximations of the phonon density of states $F(\omega)$ and the spectral electron-phonon scattering function $\alpha_{tr}^2 F(\omega)$, respectively. Both unveil a 15 meV mode, attributed to Zr vibrations in oversized B cages, which gives rise to electron-phonon coupling. The thermal expansivity further shows that this mode is anharmonic, while the vanishingly small discontinuity at $T_c$ establishes that the cell volume is nearly optimal with respect to $T_c$.

**Introduction**

The discovery of superconductivity at about 40 K in the metallic compound $MgB_2$ [1, 2] has stimulated a renewed interest in borides. The highest superconducting critical temperature in the $MB_{12}$ family is found in $ZrB_{12}$ with $T_c$ = 6 K.[3] This relatively high value compared to other dodecaborides has been considered as puzzling in view of the particularly low density of states at the Fermi level revealed by early measurements, such as the specific heat[3] and the magnetic susceptibility.[4] The exceptional hardness of this compound is reflected in the initial Debye temperature $\theta_D(0)$ of the order of 1000 K.[3] The minor fraction of transition-metal atoms appears to be responsible for superconductivity, as shown by a comparison with similar materials.[4] Measurements of the isotope effect confirm this view: the exponent $d\ln T_c / d\ln m$ for boron, −0.09±0.05,[5] is small compared to that for zirconium, −0.32±0.02.[6] The boron sublattice appears to serve as an inert background. The crystal structure (space group $Fm3m$, $a$ = 7.388 Å)[7, 8] is isotropic and based on a fcc arrangement of zirconium atoms, each of them being surrounded by 24 boron atoms arranged on a truncated octahedron. $ZrB_{12}$ has the smallest lattice constant of all known dodecaborides,[9] which suggests that the superconducting critical temperature might increase under pressure.

Recently, large and homogeneous single crystals became available,[10] and several papers addressed again the properties and the superconducting mechanism of $ZrB_{12}$. Electron transport, penetration-depth, point-contact spectroscopy (PCS), scanning tunneling spectroscopy (STS), AC and DC magnetization were investigated, with apparently conflicting results. The coupling strength appears to be weak, with a gap ratio $2\Delta(0)/k_B T_c$ = 3.64 according to transport.[11] Consistently with this result, the thermodynamic critical field obtained from magnetization measurements follows a BCS-like behavior.[12] However PCS[11] and STS[12] rather indicate very strong coupling with a gap ratio $2\Delta(0)/k_B T_c$ = 4.8. The symmetry of the superconducting wavefunction appears to be $d$-wave according to the temperature dependence of the penetration depth,[13] but $s$-wave according to STS and PCS.[11, 12] Similar discrepancies appear in the characterization of phonons. The Debye temperature is reported to be $\theta_D$ = 270-300 K according to transport,[13-15] but 930 K according to early specific-heat measurements.[3] Magnetic properties also are controversial: the upper critical field $H_{c2}(0)$ is reported to be in the range of 1100 to over 2000 G according to PCS, radio-frequency susceptibility, and transport,[11, 13] but only 390 Gauss according to DC magnetization.[12] Even the issue of type-I or type-II superconductivity is debated.[12, 13] Some of these discrepancies were discussed in terms of surface superconductivity.[12]

This situation motivated us to measure again the specific heat, a technique which, at variance with many experiments cited above, is a bulk probe sensitive to volume-averaged quantities. Results will be presented in two papers. In the present one, we report specific-heat data measured with high accuracy up to room temperature in zero magnetic field. By using an inversion process, we obtain a coarse determination of the phonon density of states (PDOS) $F(\omega)$. An important observation is that the PDOS departs markedly from the Debye model, featuring both very high cutoff energies and initial Debye temperature of the order of ~1000 K, and an excess weight at intermediate energy. In parallel with specific heat, we report resistivity data in the same temperature range. By using a similar inversion technique, we obtain a coarse picture of the spectral electron-phonon coupling function relevant for transport, $\alpha_{tr}^2 F(\omega)$. The function $\alpha_{tr}^2 F(\omega)$ is closely related to the coupling function $\alpha^2 F(\omega)$ that determines the superconducting critical temperature.[16, 17] The isotropic average of the electron-phonon matrix element $\alpha_{tr}^2 F(\omega)/F(\omega)$ is found to be particularly large at phonon energies of the order of 15 meV. Implications on superconductivity and previous apparent discrepancies are discussed.

Having measured the specific heat in the normal and in the superconducting state, we can determine the coupling strength in various ways. Our bulk data unambiguously point to weak coupling. The comparison of the Sommerfeld constant with the bare density of states at the Fermi level[18] confirms that the electron-phonon renormalization is small. We do not observe any power-law dependence of the specific heat at $T << T_c$ that could be indicative of an unconventional symmetry of Cooper pairs. Finally we address the question of the possible pressure enhancement of the critical temperature by measuring the thermal expansivity, which is thermodynamically related to the pressure dependence of $T_c$. We find that $T_c$ of $ZrB_{12}$ is nearly at its optimum with respect to the cell volume. The study of the phase diagram in the $H$-$T$ plane, showing the unusual occurrence of type-II/1 superconductivity, is presented in a second paper.[19]

# Experimental

Single crystals were grown by the Kiev group by zone melting under argon atmosphere as described in Refs[10-12], to which we refer for more detailed characterization. The large sample used for specific-heat measurements is a parallelepiped shaped by spark cutting followed by etching in a boiling $HNO_3:H_2O$ 1:1 solution. Its dimensions are $4.8 \times 4.7 \times 2.9$ mm$^3$, with the <100> axis normal to the larger face; its mass is 0.21 g. The sample used for resistivity and magnetization measurements is a rod cut parallel to the <100> direction with a diamond saw. Its length is approximately 5.3 mm, its section 0.33 mm$^2$, and its mass 6.5 mg. The demagnetization factor is $D \cong 0.03$ along the long axis.

Energy-dispersive X-ray diffraction measurements (EDX) were used to check variations of the concentration from the center to the edges of the large sample, since the surface of the original cylindrical ingot tends to be boron-deficient. The B/(B+Zr) ratio (ideally 0.923) varies significantly (i.e. beyond $2\sigma = 0.01$) only in two corners of the sample, over a layer of few tenths of a mm, where at the same time 7% at. zinc is detected. On the average, 0.5% at. silicon and 3% at. oxygen are present at the surface. These impurities are assumed to arise from spark cutting (using Cu-Zn wires), abrasive polishing, and etching. However, the volume close to the surface where at most ~10% of foreign phases is detected does not represent more than ~1% of the total volume. It follows that the global concentration of foreign phases is of the order of ~0.1%. A bulk measurement such as the specific heat is insensitive to such concentration levels, and indeed the superconducting transition shows remarkable homogeneity. However, resistance and magnetic-shielding measurements may in principle be affected. The needle-shaped sample cut with a diamond saw without further polishing nor etching is believed to be free from these problems.

The superconducting transition temperature $T_c$ was determined by four methods: resistivity (Fig. 1a), AC susceptibility (Fig. 1b), DC magnetization (Fig. 1c), and specific-heat jump at $T_c$ (Fig. 1d) which, in this order, are increasingly representative of the bulk volume (Table I). On the average, $T_c \cong 5.96$ K and the bulk transition width is ≈0.5% of $T_c$. These experiments demonstrate the remarkable bulk homogeneity of the superconducting properties of the single crystal.

The resistivity (Fig. 1a) was measured in a DC four-lead technique with offset compensation in a Quantum Design MPMS-5 magnetometer using a universal probe, a DC current source and a DC nanovoltmeter. The resistance of the contacts realized with silver paint were of the order of one Ohm each. Data were taken from 2 to 320 K in zero field, using currents of 6 and 20 mA. For this experiment, the residual field of the MPMS magnet was zeroed by monitoring the magnetization of a Pb sphere. In-field data will be reported in a second paper.[19] The low-temperature resistivity does not change appreciably below ~10 K, making the determination of the residual resistivity $\rho(0) = 1.925$ µΩ cm unambiguous. Together with $\rho(300) = 17.95$ µΩ cm at room temperature, the residual resistivity ratio is found to be $\rho(300)/\rho(0) = 9.325$. Uncertainties in the absolute values due to geometrical factors are of the order of 10%.

The Meissner magnetization (Fig. 1c) was measured in a Quantum Design MPMS-5 magnetometer using a scan length of 4 cm. The sample was cooled from above $T_c$ to 5 K in a field of 1.75 Oe (~0.4% of $H_c(0)$), followed by data acquisition during warming through the transition at a rate of ~0.01 K min$^{-1}$. Above $T_c$, the magnetization was found to be linear in $H$ up to the largest available field of 5 T. The normal-state susceptibility $\chi(T)$ is shown in Fig. 2. It is small and dominated by the diamagnetic core contribution. It does not show any spurious Curie or ferromagnetic component, at variance with early measurements.[4] The average susceptibility between 50 and 300 K is $-2.08 \times 10^{-5}$ in SI units ($-1.66 \times 10^{-6}$ uem cm$^{-3}$, $-7.75 \times 10^{-6}$ uem gat$^{-1}$). Based on band-structure calculations, the paramagnetic Pauli contribution is estimated to be $1.06 \times 10^{-5}$.[18, 20]

The specific heat was measured by a generalized relaxation technique at low temperature (1.2 - 16 K),[21] and using an adiabatic, continuous-heating calorimeter at high temperature (15 - 300 K).[21] Care was taken to zero the residual field of the 14 T magnet mounted in the cryostat. This precaution turned out to be important in view of the small critical field of ZrB$_{12}$. For this purpose, in a preliminary experiment, we maximized the superconducting transition temperature of a pure Pb wire. As a sensitive check, we verified that the superconducting transition of ZrB$_{12}$ was free from any latent heat. For normal-state specific-heat measurements, we applied a field of 5000 Oe. Results are shown in Fig. 3.

## Discussion

**Specific heat and electron-phonon coupling strength**

Normal-state specific-heat data at low temperature are analyzed in a standard way according to the formula

$$C_n(T \to 0) = \gamma_n T + \sum_{n \geq 1} \beta_{2n+1} T^{2n+1} \tag{1}$$

where the first term is the electronic contribution, with $\gamma_n = \frac{1}{3}\pi^2 k_B^2 (1+\lambda_{ep}) N_{sb}(E_F)$, $k_B$ Boltzmann's constant, $\lambda_{ep}$ the electron-phonon coupling constant, and $N_{sb}(E_F)$ the band-structure density of states including two spin directions. The second term is the low-temperature expansion of the lattice specific heat, with $\beta_3 = \frac{12}{5} N_{Av} k_B \pi^4 \theta_D^{-3}(0)$, with $N_{Av}$ Avogadro's number, and $\theta_D(0)$ the initial Debye temperature. From a fit of normal-state data from 1.2 to 16 K, we find $\gamma_n$ = 0.34 mJ K$^{-2}$ gat$^{-1}$ and $\theta_D(0)$ = 970 K. The Sommerfeld constant $\gamma_n$ corresponds to a dressed density of states at the Fermi level $N(E_F) = 0.144$ states eV$^{-1}$ atom$^{-1}$. Compared to recent band-structure calculations,[18, 20] there is room for a small renormalization factor $1 + \lambda_{ep} \cong 1.2$; in other words ZrB$_{12}$ is in the weak-coupling regime of superconductivity. Consistently with this conclusion, we find that the normalized specific-heat jump is $\Delta C / \gamma_n T_c$ = 1.66, not far from the weak-coupling BCS limit 1.43.

The thermodynamic critical field $H_c(T)$ is obtained by numerical integration of the data:

$$-\frac{1}{2}\mu_0 V H_c^2(T) = \Delta F(T) = \Delta U(T) - T\Delta S(T),$$

$$\Delta U(T) = \int_T^{T_c} [C_s(T') - C_n(T')]dT',$$

$$\Delta S(T) = \int_T^{T_c} \frac{C_s(T') - C_n(T')}{T'} dT', \tag{2}$$

where $V$ is the volume of one gram-atom (1 gat = 1/13 mol corresponds to 4.68 cm$^3$ or 17.0 g) and the specific heat $C$ is given per gat. We obtain $H_c(0) = 415$ Oe at $T = 0$. The deviation function $D(t) \equiv h - (1 - t^2)$, where $h \equiv H_c(T)/H_c(0)$ and $t \equiv T/T_c$, is comparable with that of In or Sn,[22, 23] and again characteristic of weak coupling. A more quantitative description of the same experimental information is given by the α-model.[24] We find that the supposedly constant ratio $\Delta_{ZrB_{12}}(T)/\Delta_{BCS}(T)$ that best fits the normalized electronic specific-heat curve over the whole temperature range is 1.05, i.e. $2\Delta_{ZrB_{12}}(0)/k_B T_c = 3.7$, again only slightly above the weak-coupling limit $2\Delta_{BCS}(0)/k_B T_c = 3.53$. In another approach, we make use of the strong-coupling corrections calculated within the isotropic single-band model on the basis of Eliashberg theory as given by Eq. (14) and (15) of Ref[25]. Taking the measured specific-heat jump ratio $\Delta C/\gamma_n T_c = 1.66$ as an input parameter, we obtain for the characteristic logarithmic average energy $\omega_{ln} \equiv \exp\left(\int \ln \omega \alpha^2 F(\omega) d\omega/\omega \middle/ \int \alpha^2 F(\omega) d\omega/\omega\right)$ = 159 K, and $2\Delta_{ZrB_{12}}(0)/k_B T_c$ = 3.69, in perfect agreement with the result of the α-model. It is concluded that ZrB$_{12}$ definitely is a weak-coupling superconductor. Bulk values of the gap ratio obtained here and values as high as 4.8 obtained by spectroscopic methods[11, 12] are mutually exclusive. They can only be reconciled if the gap is enhanced at the surface, rather than degraded as would generally be the case for chemical inhomogeneity. This puzzling issue would deserve additional spectroscopic investigations.

We finally note that the quality of the α-model fit and the shape of the deviation function exclude $d$-wave superconductivity. In particular, in the latter case, the dimensionless ratio $\gamma_n T_c^2/\mu_0 V H_c^2(0)$ would be nearly twice as large (3.7) as that measured (1.9).[26]

**Specific heat and phonon density of states**

The Debye temperature $\theta_D(0) = 970$ K given by the low-temperature expansion of the specific heat is consistent with measurements of the ultrasonic sound velocity, $\theta_D^{elast}$ = 1040 K.[27] However, the Debye temperature is not constant. Fig. 4 shows ideal Debye curves $C_D(T)$ for selected values of $\theta_D$, together with the lattice specific heat measured up to room temperature. It is clear than the data do not follow any curve with a constant $\theta_D$, and in particular are inconsistent with $\theta_D \cong 270\text{-}300$ K obtained from fits of the resistivity given in Refs[13-15]. The effective Debye temperature is defined as the value $\theta_D(T)$ such that the tabulated ideal Debye specific heat[28] $C_D(\theta_D/T)$ equals the measured lattice specific heat $C_{ph}(T) = C(T) - \gamma_n T$ at a given temperature $T$. A correction for the temperature dependence of $\gamma_n$ does not appear necessary in view of the absence of structure in the density of states near the Fermi level[18] and the small electron-phonon renormalization.[16] The effective Debye temperature is plotted in the inset of Fig. 4. Main features are: (i) $\theta_D$ at room temperature, 1200 K, is close to $\theta_D$ at 10 K; (ii) there is deep minimum at $T_m \approx 35$ K. Whereas point (i) means that both the initial curvature of $F(\omega)$ and the cutoff frequency $\omega_D$ are apparently consistent with a single Debye component, point (ii) reflects additional weight at intermediate energies $\omega_E \approx 5T_m \approx 15$ meV. Analogous situations occur e.g. in Na and Al due to the presence of optical phonons[28] and in hexaborides.[29]

The specific-heat data at high temperature have sufficiently low scatter (~0.02%) to attempt an inversion of $C_{ph}(T)$ to extract the PDOS $F(\omega)$. More precisely, the problem being generally ill-conditioned, we can only obtain a substitutional spectrum, i.e. a smoothed phonon density of states $\overline{F}(\omega)$ that reproduces precisely the specific heat and low-order moments of $F(\omega)$, but may not show the true PDOS in detail. A simplified method consists in representing $F(\omega)$ by a geometrical series of Einstein modes with fixed positions and adjustable weights:

$$F(\omega) = \sum_k F_k \delta(\omega - \omega_k), \tag{3}$$

The corresponding lattice specific heat is given by:

$$C(T) = 3R \sum_k F_k \frac{x_k^2 e^{x_k}}{(e^{x_k} - 1)^2}, \tag{4}$$

where $x_k = \omega_k / T$. The weights $F_k$, normalized to a total of one if the specific heat is given per gram-atom, are found by a least-squares fit of the measured specific heat. The number of modes is chosen to be small enough so that the solution is stable; in practice a good choice is $\omega_{k+1}/\omega_k = 1.75$. Figure 5 shows the decomposition of the lattice specific heat into Einstein contributions. The PDOS obtained in this way is shown in Fig. 6. Qualitatively similar to that of $LaB_6$,[29] it consists of a quasi-Debye background with a high characteristic frequency, as expected in view of the boron mass, superposed onto an optical mode with energy ~15 meV, presumably associated with the oscillations of the Zr atoms in the boron "cages" present in the structure. The relative weight of the latter mode, 5.3%, is not far from the fraction of Zr atoms per formula unit, 1/13; part of the missing weight may be located in the neighboring energy bins. The question arises, whether the latter mode strongly contributes to the electron-phonon coupling. This point is addressed in the next section, using resistivity as an experimental probe.

**Resistivity and electron-phonon coupling**

The resistivity (Fig. 7) is analyzed in a similar way. We start from the generalized Bloch-Grüneisen formula (see e.g. Ref[16], in particular p. 212 and 219):

$$\rho(T) = \rho(0) + \frac{4\pi m}{ne^2} \int_0^{\omega_{max}} \alpha_{tr}^2 F(\omega) \frac{xe^x}{(e^x - 1)^2} d\omega \qquad (5)$$

where $x \equiv \omega/T$ and $\alpha_{tr}^2 F(\omega)$ is the electron-phonon "transport coupling function". In the restricted Bloch-Grüneisen approach, one would have $\alpha_{tr}^2 F(\omega) \propto \omega^4$, and as a consequence $\rho(T) - \rho(0) \propto T^5$, but deviations from the Debye model, complications with phonon polarizations, and Umklapp processes would not justify this simplification beyond the low-temperature continuum limit. Using a decomposition into Einstein modes similar to Eq. (3),

$$\alpha_{tr}^2 F(\omega) = \sum_k \alpha_k^2 F_k \delta(\omega - \omega_k), \qquad (6)$$

we obtain the discrete version of Eq. (5):

$$\rho(T) = \rho(0) + \frac{4\pi m}{ne^2} \sum_k \alpha_k^2 F_k \frac{x_k e^{x_k}}{(e^{x_k}-1)^2} \qquad (7)$$

where the fitting parameters are the coefficients $\alpha_k^2 F_k$ and $\rho(0)$. Fig. 8 shows the decomposition of the total resistivity into Einstein components. In agreement with specific heat, the main component arises from modes with energies near 15 meV. In the absence of information on $n/m$, the amplitudes of $\alpha_{tr}^2 F(\omega)$ shown in Fig. 6 are normalized arbitrarily. Two independent runs performed with different sets of contacts were analyzed separately; their results shown in Fig. 6 demonstrate the stability of the inversion at low energy.

The electron-phonon transport coupling function $\alpha_{tr}^2 F(\omega)$ is closely related to the function $\alpha^2 F(\omega)$ which governs superconductivity.[17] Compared to other modes, the ~170 K region is weighted much more heavily in the $\alpha_{tr}^2 F(\omega)$ function than in the PDOS $F(\omega)$. Together with isotope-effect experiments showing the importance of the zirconium mass, these observations suggest that most of the electron-phonon coupling arises from the low frequency vibration (~15 meV) of loosely bound zirconium atoms in their boron cages. It is interesting to compare this situation with that of $MgB_2$. In the latter case, most of the coupling comes from the $E_{2g}$ mode having an energy of 700 to 900 K.[30] Remembering that the phonon energy appears as a prefactor in $T_c$ equations,[17] this difference by a factor of 4 to 5 explains much of the difference between $T_c$ of $ZrB_{12}$ (6 K) and that of $MgB_2$ (39 K). The low dimensionality and multiple gap structure of $MgB_2$ may be invoked to explain the rest. This difference also explains why the resistivity of $ZrB_{12}$ behaves so differently from that of $MgB_2$.[15]

Another remark confirms that the ~15 meV mode rather than the whole spectrum determines superconductivity in $ZrB_{12}$. Based on strong-coupling corrections to the specific heat jump, we have found that the characteristic energy of phonons which appears as a prefactor in Allen and Dynes' $T_c$ equation[17] is $\omega_{\ln} \equiv \exp\left(\int \ln\omega \, \alpha^2 F(\omega) d\omega/\omega \Big/ \int \alpha^2 F(\omega) d\omega/\omega\right) \cong 14$ meV. This is close to the energy of the low-lying mode, and far from the uniformly weighted logarithmic average $\exp\left(\int \ln\omega F(\omega) d\omega/\omega \Big/ \int F(\omega) d\omega/\omega\right) = 42$ meV.

The generalized Bloch-Grüneisen equation used here fits the measured resistivity over the whole temperature range from 2 to 300 K, without requiring any additional electron-electron scattering term $\rho_{e-e} \propto T^2$. Different conclusions were obtained in the literature by assuming that the Debye temperature has a value between 270 and 300 K and remains constant at least up to 25 K.[13-15] The latter assumption is obviously not supported by specific-heat measurements (Fig. 4); the initial Debye temperature is close to 1000 K and the continuum approximation $F(\omega) \propto \omega^2$ does not hold beyond 10 K. At 25 K the deviation with respect to the initial $C_{phonon}(T) \propto T^3$ law is as high as 300%. Therefore the full structure of the PDOS must be taken into account to understand both the specific heat and the resistivity. This is further confirmed by thermal-expansion measurements, as shown below.

**Thermal expansivity and anharmonicity**

Thermal-expansion experiments were undertaken to give three types of information: (i) confirm the main features of the PDOS; (ii) evaluate the volume dependence of phonon modes; (iii) determine the variation of $T_c$ with pressure. The linear thermal expansivity $\alpha(T)$ for a cubic system is given by:

$$\alpha(T) \equiv \frac{1}{L}\left(\frac{\partial L}{\partial T}\right)_p = \frac{\kappa_T}{3}\left(\frac{\partial S}{\partial V}\right)_T, \qquad (8)$$

where $\kappa_T$ is the isothermal compressibility. The expansivity, which is closely related to the lattice specific heat at constant volume via the Grüneisen parameter $\gamma_G \equiv 3\alpha V/\kappa_T C_v$, actually measures the PDOS with a weighting $\gamma_G(\omega) \equiv -\partial \ln\omega/\partial \ln V$ that represents anharmonicity. Measurements were performed in Karlsruhe in a spring-loaded cell with capacitive detection.[31, 32] Results are shown in Fig. 9. The nearly linear expansivity in the temperature range from 100 to 300 K is typical for a high Debye temperature. Around 50 K a broad anomaly appears, which is an indication for low-lying modes with a strong anharmonicity. A simple fit is obtained using a contribution from an Einstein mode ($T_E^\alpha = 170$ K) and another one from a Debye distribution ($\theta_D^\alpha = 1400$ K), as shown by the dotted and dashed lines, respectively. While $T_E^\alpha$ and $\theta_D^\alpha$ are consistent with the main features of the spectra determined by specific heat and resistivity, the considerable weighting of the $T_E^\alpha$ component in the expansivity evidences large volume dependence for this mode. This is also apparent in the behavior of the average Grüneisen parameter $\gamma_G(T)$ (Fig.10), evaluated assuming $\kappa_T^{-1}$ = 248 GPa by analogy with $UB_{12}$.[33] The signature of the 15meV modes occurs near 35 K ($\approx T_E^\alpha/5$) in this plot, as was the case for $\theta_D(T)$. For a more quantitative analysis we performed a similar fit for the thermal expansivity as for the specific heat based on a series of Einstein functions using the same frequencies. Equation 9 allows us to extract the Grüneisen parameters $\gamma_{G_k}$ for each phonon frequency separately (inset of Fig.10):

$$\alpha(T) = \frac{3R\kappa_T}{3V} \sum_k \gamma_{G_k} F_k \frac{x_k^2 e^{x_k}}{(e^{x_k}-1)^2}, \qquad (9)$$

where the $F_k$ are taken from Eq. 4. The 170-K mode is heavily weighted with $\gamma_{G_k}$ = 4.1, whereas the other modes are much less anharmonic with $\gamma_{G_k}$ values below 2. Similarly to $MgB_2$, the anharmonic mode is the one that gives rise to a large electron-phonon coupling.

The pressure dependence of $T_c$ is obtained from the Ehrenfest relation

$$\Delta\alpha = \frac{1}{3V} \frac{\Delta C}{T_c} \left(\frac{\partial T_c}{\partial p}\right)_T, \qquad (10)$$

where $\Delta$ stands for discontinuities at the second-order transition. No such discontinuity can be observed in $\alpha$ at 6 K within the experimental scatter of $\sim 10^{-8}$ K$^{-1}$ (inset of Fig. 9). This sets an upper limit of 0.24 K/GPa for the absolute value of the pressure dependence of $T_c$. As this value is rather small[34], one can conclude that the cell volume of $ZrB_{12}$ is close to its optimal value for superconductivity. This means that the increase of phonon frequencies with pressure which, taken alone, should raise $T_c$, is compensated by changes in the electronic structure.

## Conclusion

Specific-heat, resistivity and thermal-expansion experiments performed on high-quality single crystals have been used to characterize $ZrB_{12}$ (Table II). At variance with some published data, we find that this superconductor, which has a low density of states at the Fermi level, lies in the weak-coupling regime, as it shows all typical characteristics of single-band, isotropic BCS superconductivity. The apparent discrepancy between various estimations of the Debye temperature is resolved by showing that the PDOS of $ZrB_{12}$ contains both very high frequencies ~100 meV and a high density of low-lying modes near 15 meV. The comparative analysis of specific-heat, resistivity and thermal-expansion data leads to the conclusion that the latter low-lying modes, which are anharmonic and coupled to the electronic system, play an important role in the superconductivity of $ZrB_{12}$. They are tentatively associated with the vibration of zirconium atoms loosely bound in the oversized cages formed by boron ions, a situation analogous to that of $LaB_6$ described by Mandrus et al.[29] One distinct characteristics of $ZrB_{12}$ compared to $MgB_2$ is the low frequency of the phonon modes which are coupled to electrons. We have shown that this cannot be remedied by decreasing the cell volume. Substituting Zr by a lighter element should favour higher $T_c$. The unusual magnetic properties of the superconducting state of $ZrB_{12}$ will be addressed in a second paper.[19]

## Acknowledgements

Stimulating discussions with V. Gasparov, J. Kortus, I.R. Shein, J. Teyssier, and T. Jarlborg are gratefully acknowledged; in particular V. Gasparov attracted our attention to $ZrB_{12}$ and passed on samples. We thank M. Schindl for expertise in EDX measurements, and A. Naula for technical help. This work was supported by the National Science Foundation through the National Centre of Competence in Research "Materials with Novel Electronic Properties–MaNEP".

# Tables

Table I. Critical temperature and transition width measured by different methods. Differences result from independent temperature calibrations, differences in heating rates and data density.

|  | $T_c$ midpoint (K) | $\Delta T_c$ (mK) |
|---|---|---|
| Resistivity | 5.97 | <25 (0-100%) |
| AC susceptibility @ 8 kHz, 0.01 Oe | 5.98 | 80 (10-90%) |
| Meissner magnetization @ 1.75 Oe | 6.00 | 20 (10-90%) |
| Specific-heat jump | 5.91 | <30 (0-100%) |

Table II. Characteristic parameters of ZrB$_{12}$. $T_c$, superconducting transition temperature; $RRR$, residual resistivity ratio; $V$ and $M$, mean atomic volume and mass, respectively; $\gamma_n$, Sommerfeld constant; $\Delta C/\gamma_n T_c$, normalized specific-heat jump at $T_c$; $\Theta_D(0)$, Debye temperature at $T \to 0$; $H_c(0)$, thermodynamic critical field at $T \to 0$; $2\Delta(0)/k_B T_c$, normalized superconducting gap; $\chi$, normal-state magnetic susceptibility; $N_{sb}(E_F)$, bare density of states at the Fermi level (per 13-atom quarter-cell).[20]

| | |
|---|---|
| $T_c$ (K) | $5.96 \pm 0.05$ |
| $RRR$ | $9.33 \pm 0.03$ |
| $V$ (cm$^3$ gat$^{-1}$) | 4.68 |
| $M$ (g gat$^{-1}$) | 17.0 |
| $\gamma_n$ (mJ K$^{-2}$ gat$^{-1}$) | $0.34 \pm 0.02$ |
| $\Delta C/\gamma_n T_c$ | $1.66 \pm 0.1$ |
| $\theta_D(0)$ (K) | $970 \pm 20$ |
| $H_c(0)$ (mT) | $41.5 \pm 1$ |
| $2\Delta(0)/k_B T_c$ | $3.7 \pm 0.1$ |
| $\gamma_n T_c^2 / \mu_0 V H_c^2(0)$ | $1.88 \pm 0.2$ |
| $\chi$ (S.I.) | $-2.08 \times 10^{-5}$ |
| $N_{sb}(E_F)$ (eV cell)$^{-1}$ | 1.59 |

**Figure captions**

Fig. 1. Superconducting transition of the ZrB$_{12}$ crystal observed by (a) resistivity, (b) AC susceptibility, (c) Meissner magnetization, and (d) specific heat.

Fig. 2. Magnetic susceptibility in the normal state as a function of the temperature. Dashed line: average between 50 and 300 K.

Fig. 3. Total specific heat $C/T$ of ZrB$_{12}$ in the superconducting state (open symbols) and in the normal state (closed symbols) versus temperature squared. The dashed lines below and above the specific-heat peak at $T_c$ are calculated using the α-model with $2\Delta(0)/k_B T_c = 3.5$ and 3.9, respectively (best fit: 3.7). Inset: specific heat $C$ versus temperature up to room temperature (~3000 independent data).

Fig. 4. Lattice specific heat of ZrB$_{12}$ versus temperature up to 300 K. The dashed lines show, from left to right, the ideal Debye specific heat calculated for $\theta_D = 270$, 635, 865, and 1200 K. Inset: effective Debye temperature versus temperature.

Fig. 5. Lattice specific heat divided by the cube of the temperature showing the decomposition into Einstein terms. The main peak at ~35 K is associated with the Einstein mode at $\omega_E = 170$ K (15 meV).

Fig. 6. Full line, closed symbols: phonon density of states $F(\omega)$ obtained by inversion of the specific heat. Dashed lines, open symbols: spectral electron-phonon transport coupling function $\alpha_{tr}^2 F(\omega)$ obtained by inversion of the resistivity, showing the results of two independent runs.

Fig. 7. Resistivity of ZrB$_{12}$ versus temperature. Main frame: zero magnetic field. Inset: expanded low-temperature normal-state region in a field of 1 T.

Fig. 8. Total resistivity divided by the temperature showing the decomposition into Einstein terms and the residual term. The largest Einstein component is centered on $\omega = 170$ K. Residuals $R/R_{fit} - 1$ are shown by crosses.

Fig. 9. Linear thermal expansivity of ZrB$_{12}$ versus temperature. The dotted and dashed lines show the decomposition into Einstein and Debye components, respectively; the resulting fit is hidden by the data. Inset: expanded view near the superconducting transition at 6 K.

Fig. 10. Grüneisen parameter of ZrB$_{12}$ versus temperature, assuming a bulk modulus $\kappa_T^{-1} = 248$ GPa. Inset: open diamonds, Grüneisen parameter of each phonon modes

obtained from the fit using Eq. 9; full circles, the phonon density of states $F(\omega)$ as plotted in Fig.6.

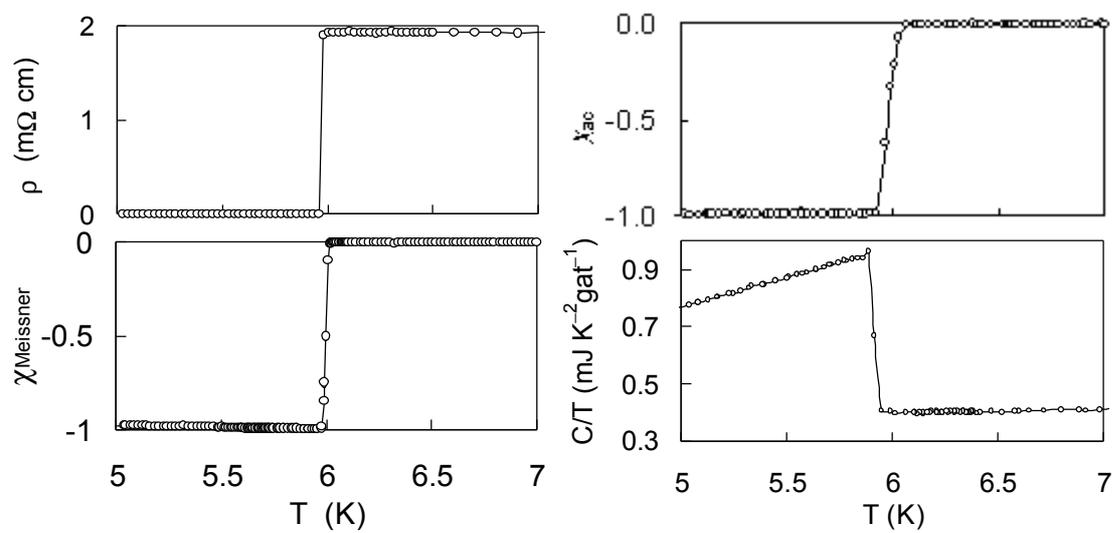

1a 1b
1c 1d

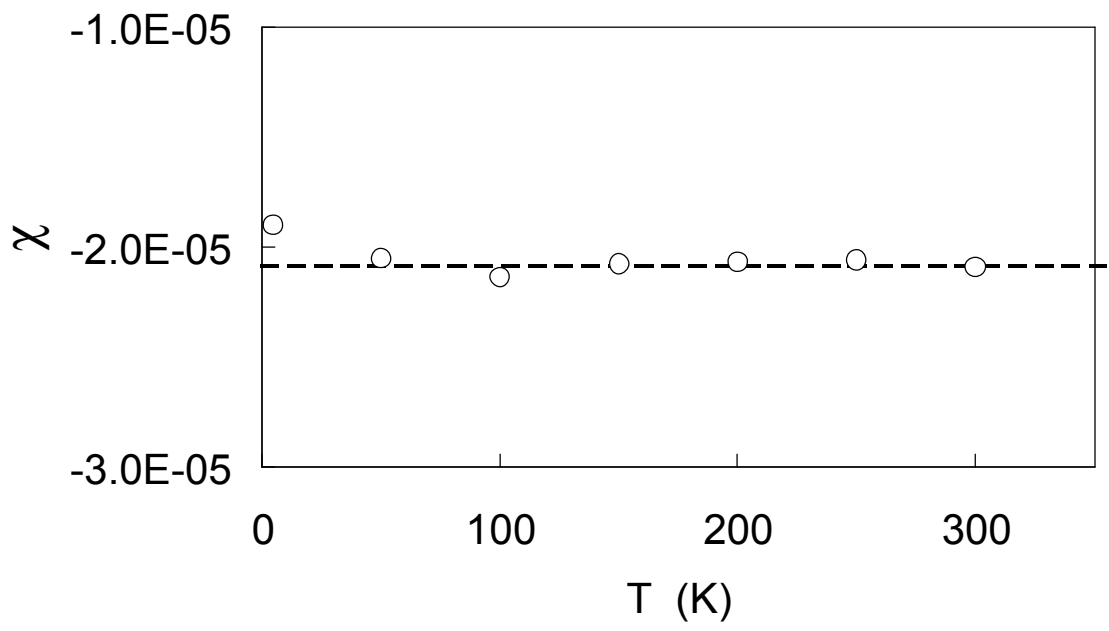

**2**

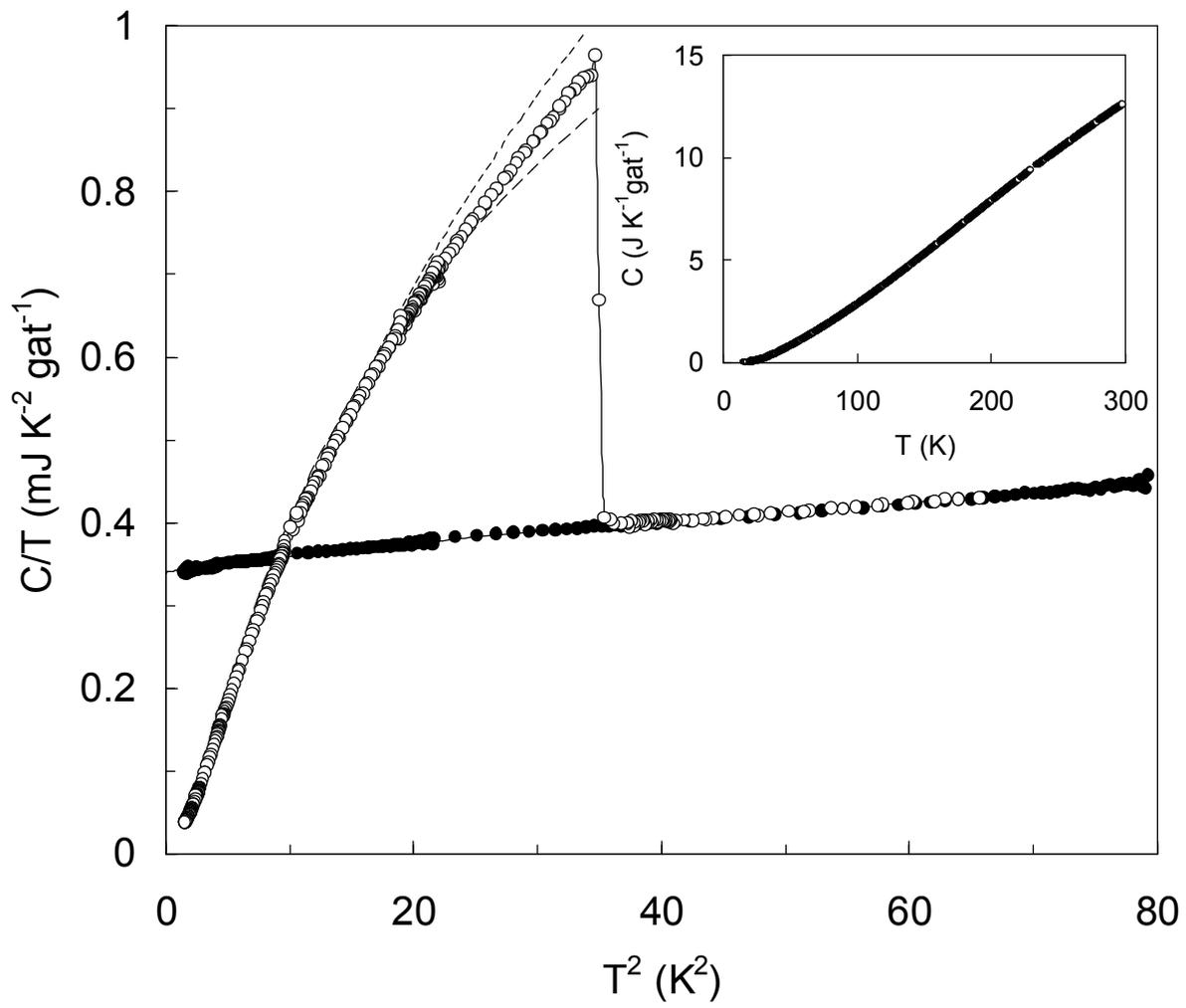

**3**

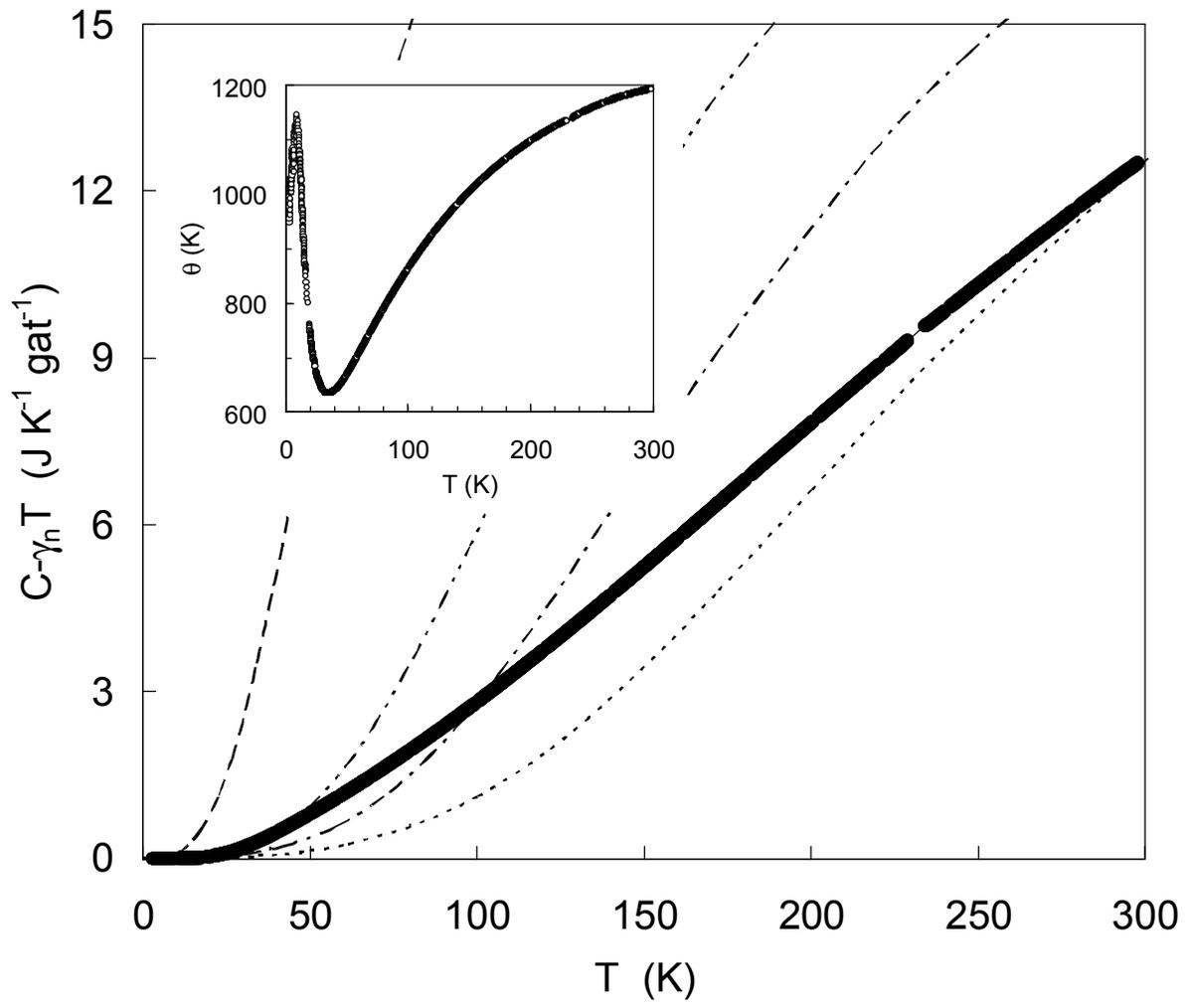

**4**

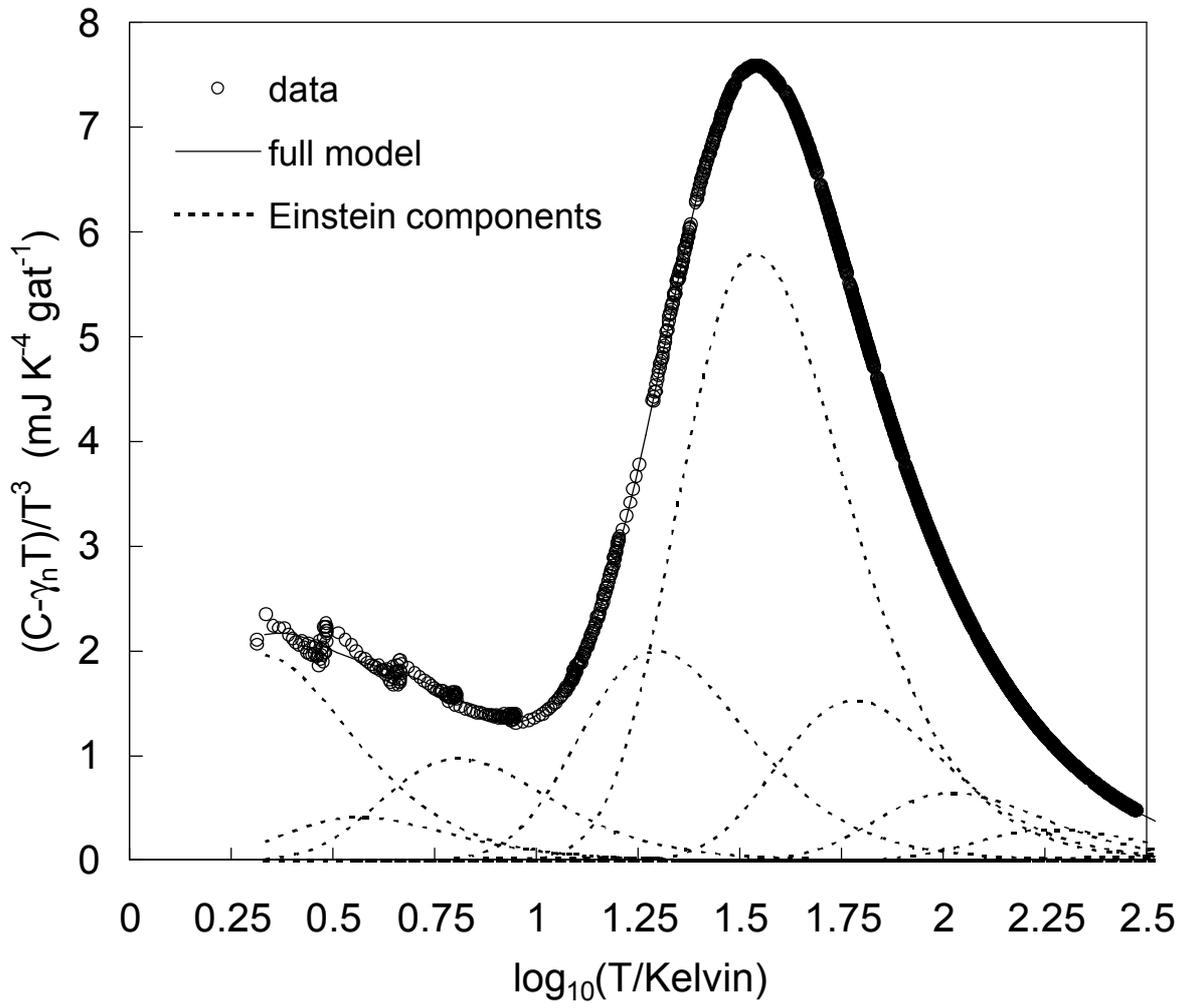

**5**

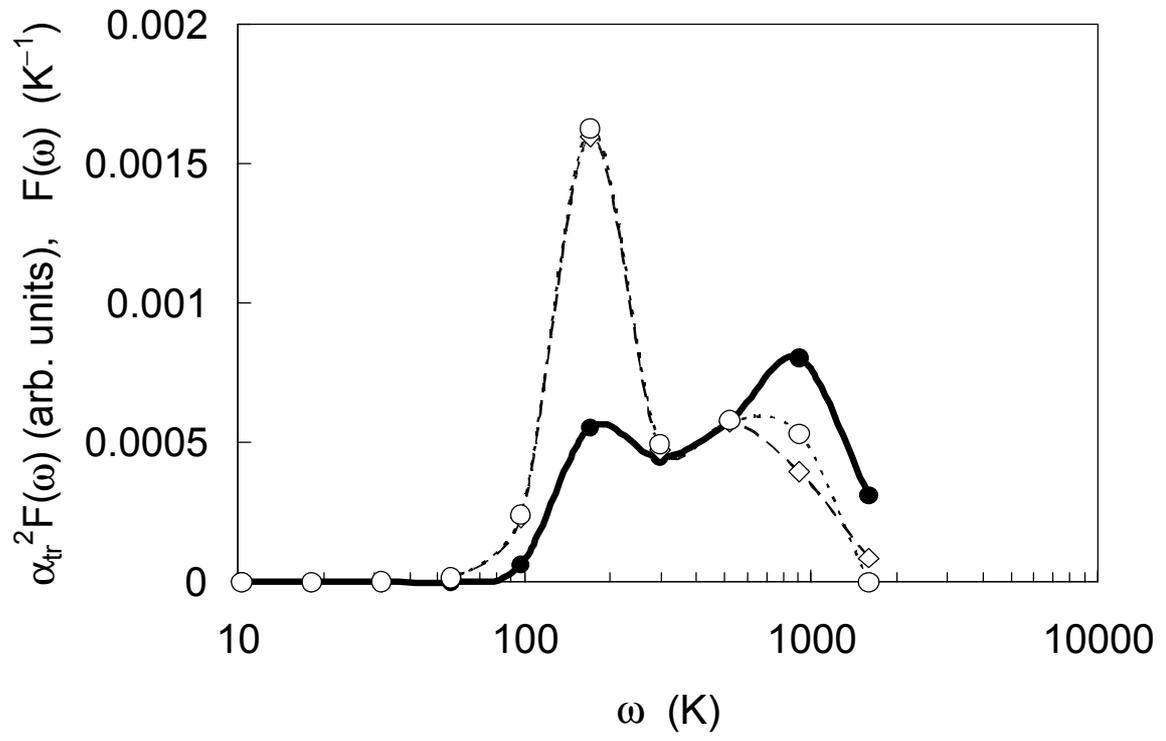

**6**

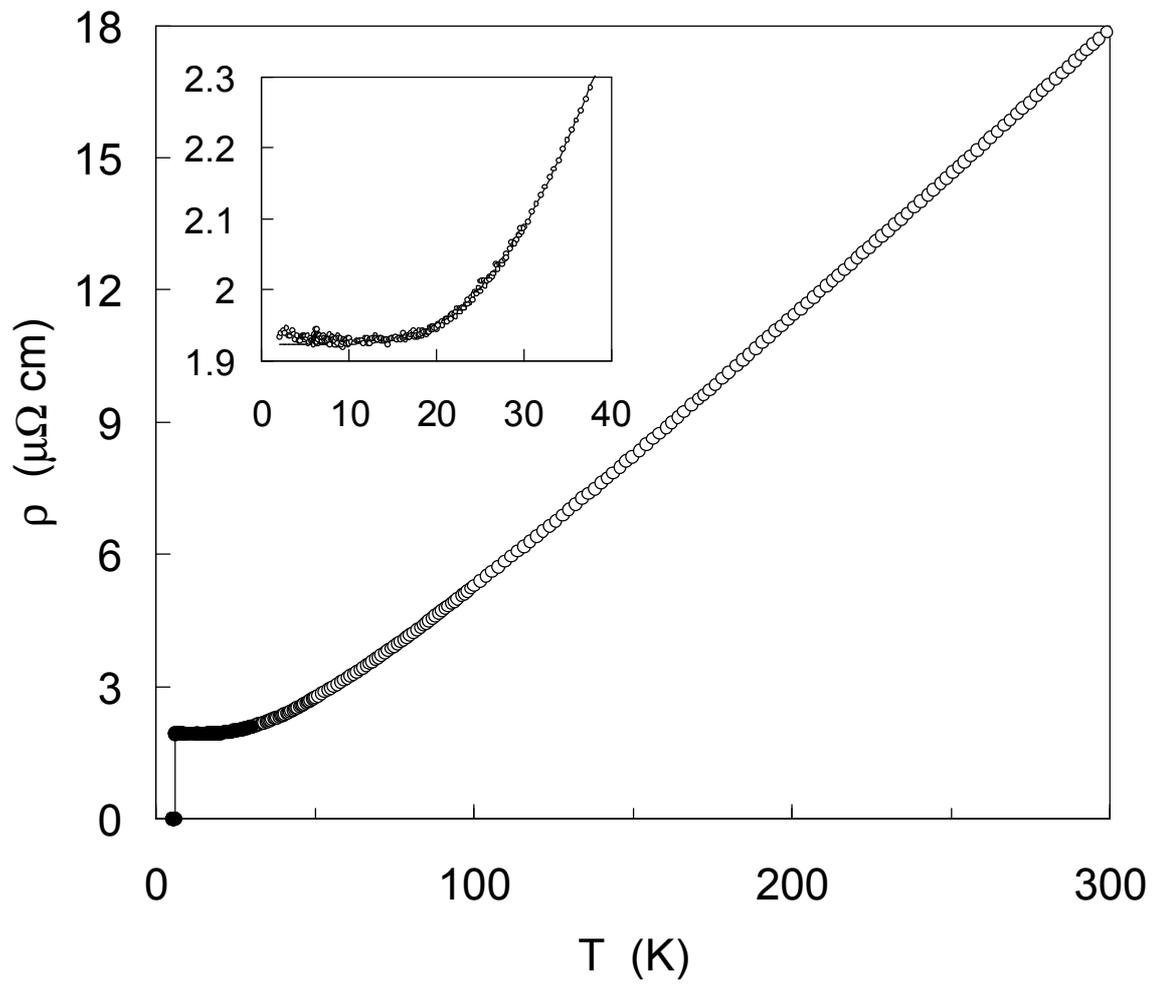

**7**

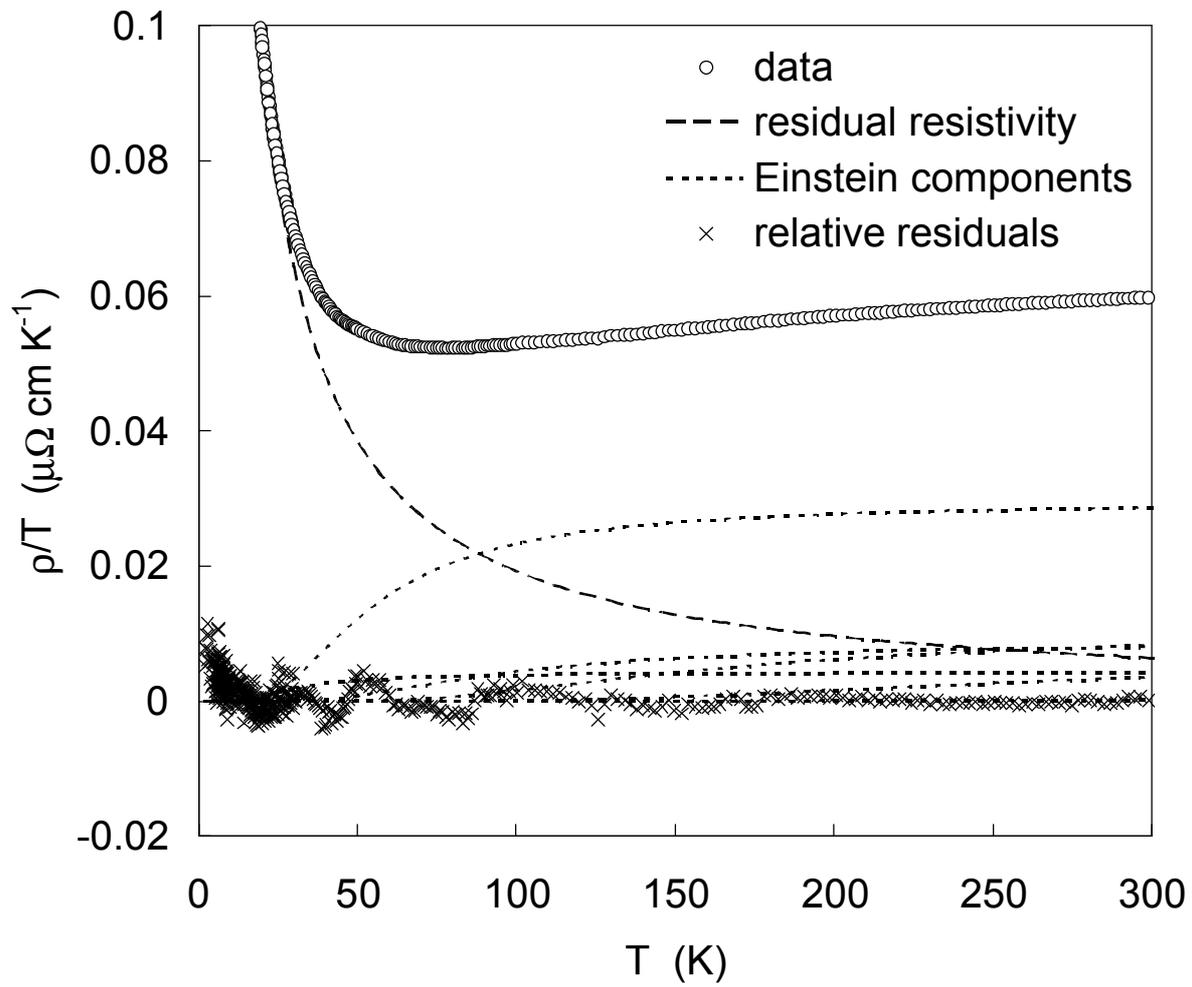

**8**

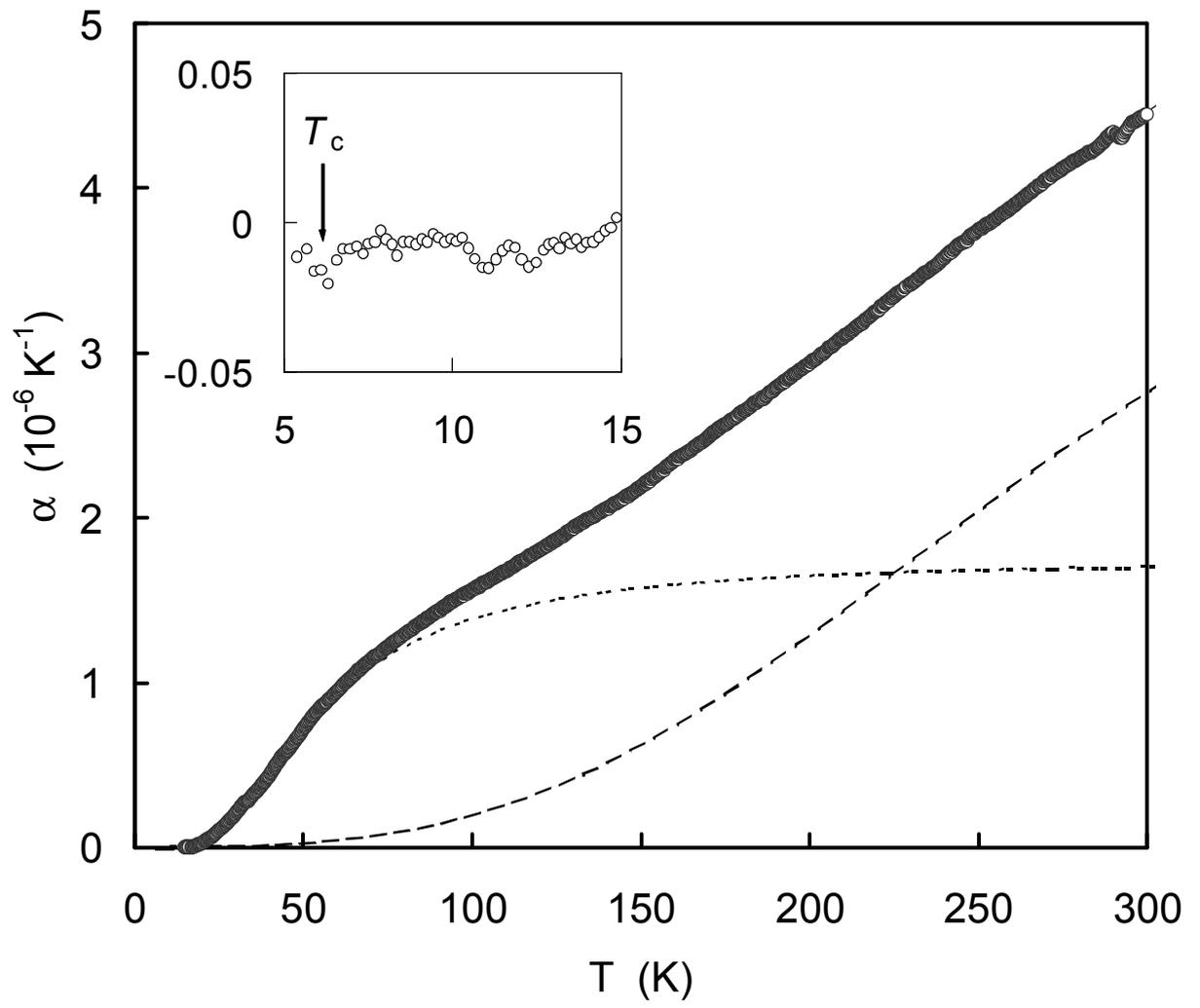

**9**

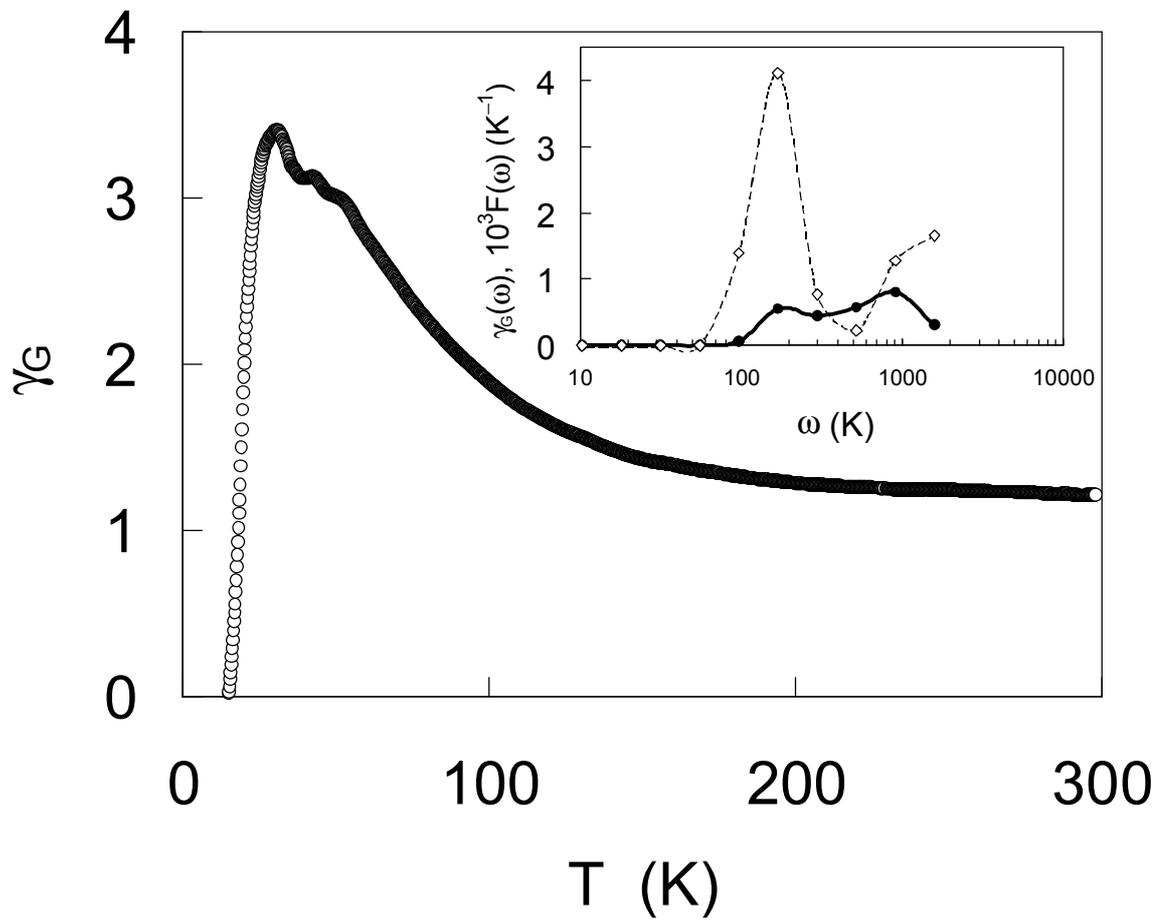